\def \SAIT #1 #2 {{\em Mem.\ Soc.\ Astron.\ It.\/} {\bf #1}, #2}
\def \MESS #1 #2 {{\em The Messenger\/} {\bf #1}, #2}
\def \ASTRNACH #1 #2 {{\em Astron. Nach.\/} {\bf #1}, #2}
\def \AAP #1 #2 {{\em Astron. Astrophys.\/} {\bf #1}, #2}
\def \AAL #1 #2 {{\em Astron. Astrophys. Lett.\/} {\bf #1}, L#2}
\def \AAR #1 #2 {{\em Astron. Astrophys. Rev.\/} {\bf #1}, #2}
\def \AAS #1 #2 {{\em Astron. Astrophys. Suppl. Ser.\/} {\bf #1}, #2}
\def \AJ #1 #2 {{\em Astron. J.\/} {\bf #1}, #2}
\def \ANNREV #1 #2 {{\em Ann. Rev. Astron. Astrophys.\/} {\bf #1}, #2}
\def \APJ #1 #2 {{\em Astrophys. J.\/} {\bf #1}, #2}
\def \APJL #1 #2 {{\em Astrophys. J. Lett.\/} {\bf #1}, L#2}
\def \APJS #1 #2 {{\em Astrophys. J. Suppl.\/} {\bf #1}, #2}
\def \APSS #1 #2 {{\em Astrophys. Space Sci.\/} {\bf #1}, #2}
\def \ASR #1 #2 {{\em Adv. Space Res.\/} {\bf #1}, #2}
\def \BAIC #1 #2 {{\em Bull. Astron. Inst. Czechosl.\/} {\bf #1}, #2}
\def \JSQRT #1 #2 {{\em J. Quant. Spectrosc. Radiat. Transfer\/} {\bf #1}, #2}
\def \MN #1 #2 {{\em Mon. Not. R. Astr. Soc.\/} {\bf #1}, #2}
\def \MEM #1 #2 {{\em Mem. R. Astr. Soc.\/} {\bf #1}, #2}
\def \PLR #1 #2 {{\em Phys. Lett. Rev.\/} {\bf #1}, #2}
\def \PASJ #1 #2 {{\em Publ. Astron. Soc. Japan\/} {\bf #1}, #2}
\def \PASP #1 #2 {{\em Publ. Astr. Soc. Pacific\/} {\bf #1}, #2}
\def \NAT #1 #2 {{\em Nature\/} {\bf #1}, #2}
\documentstyle[twoside,epsf]{memsait}

\begin{opening}
\title{s- and r-process elements in two very metal-poor stars}
\author{
Sean G. Ryan$^1$, Wako Aoki$^2$, Lisa A. J. Blake$^1$, John E. Norris$^3$,
Timothy C. Beers$^4$, Roberto Gallino$^5$, Maurizio Busso$^6$, and 
Hiroyasu Ando$^2$
}
\institute{
$^1$Dept of Physics \& Astronomy, The Open University, Walton Hall, Milton 
Keynes MK7~6AA, United Kingdom; s.g.ryan@open.ac.uk, l.a.j.blake@open.ac.uk
$^2$National Astronomical Observatory, Mitaka, Tokyo, 181-8588 Japan;
aoki.wako@nao.ac.jp, ando@optik.mtk.nao.ac.jp
$^3$Research School of Astronomy and Astrophysics, The Australian National
University, Private Bag, Weston Creek Post Office, Canberra, ACT 2611, 
Australia; jen@mso.anu.edu.au
$^4$Department of Physics and Astronomy, Michigan State University, E. Lansing,
MI 48824-1116, USA; beers@pa.ms.edu
$^5$Dipartimento di Fisica Generale, Universit\'a\ di Torino, Via P. Giuria 1, 10125 Torino, Italy; gallino@ph.unito.it
$^6$Osservatorio Astronomico di Torino, 10025 Pino Torinese, Italy; busso@to.astro.it
}
\date{} % DO NOT INSERT ANY DATE HERE !!!
\end{opening}

\begin{document}

%\oddpagefooter{\sf Mem. S.A.It., Vol. ??, 1994}{}{\thepage}
%\evenpagefooter{\thepage}{}{\sf Mem. S.A.It., Vol. ??, 1994}
\oddpagefooter{}{}{} % LEAVE AS IT IS !
\evenpagefooter{}{}{} % LEAVE AS IT IS !
\ 
\bigskip

\begin{abstract}
New measurements of neutron-capture elements are presented for two very
metal-poor stars ([Fe/H] $\sim$ $-3$). One (LP~625-44) has an s-process 
signature believed to be due to mass transfer from a now-extinct metal-poor AGB 
companion, and the second (CS~22897-008) is one of a number of very metal-poor 
stars having high [Sr/Ba] ratios which is not expected from the r-process.
In the s-process star, many elements including lead have been detected,
providing strong constraints on the $^{13}$C pocket in the now-extinct AGB star.
In the Sr-rich star, Zn, Y, and Zr are also seen to be overabundant, and 
several possible nucleosynthesis mechanisms are discussed.
\end{abstract}

\section{Introduction}

Neutron-capture elements observed in Population II stars are generally
interpreted in a framework developed some 20 years ago. Spite \& Spite (1978)
showed from observations of Ba (a mixture of 80\%\ s- and 20\%\ r-process 
fractions in the solar system; e.g., Arlandini et al. 1999) and Eu 
(primarily r-process) that Pop II stars exhibit
r-process abundance patterns. Truran (1981) provided
a theoretical basis for this result by noting that the s-process needed
pre-stellar seed nuclei, whereas an r-process site would 
produce its own seed nuclei. For this reason, metal-deficient populations
would be poor producers of s-process nuclei. It became common to regard
neutron-capture elements in Pop II stars as due solely to the r-process.
Observational support for this view grew with the work of Gilroy et al. (1988)
and McWilliam et al. (1995).
The theory is demonstrated numerically in the Galactic chemical 
evolution (GCE) calculations of 
Travaglio et al. (1999; see also Busso, Gallino \& Wasserburg 1999, Fig. 17)
which showed that s-process products do not feature significantly in 
newly forming stars until [Fe/H] $^>_\sim$ $-1.5$.  
(See also Raiteri et al. 1999). 

Observations of several metal-poor stars 
enriched in neutron-capture elements have revealed solar
r-process patterns. That is, despite their considerable metal deficiency, 
these stars seem to have experienced an r-process that barely differs from
the sum of processes that enriched the pre-solar nebula. This 
has led to suggestions that r-process production may be independent
of the initial metallicity of the site and, to some degree, universal throughout
Galactic history
(e.g. Cowan et al. 1995; Sneden et al. 1996).
Such a finding, if widely substantiated, would have important 
implications for identifying the r-process site.
Theoretical work in the last 
decade has focussed on the neutrino-heated bubble located above the surface of a
newly-formed neutron star (e.g. Woosley et al. 1994;
Takahashi, Witti, \& Janka 1994), but other sites such as colliding
neutron stars may provide an alternative, at least for the heavier r-process
nuclei with atomic mass $A > 130$ 
(e.g. Freiburghaus, Rosswog, \& Thielemann 1999).

Solar r-process abundance ratios of Ba and Eu are seen in many Pop II 
stars. In a small number of stars including CS~22892-052 
(Cowan et al. 1995; Sneden et al. 1996), an even larger range of elements can be
measured and likewise conform to that pattern. However, not all neutron-capture 
elements are present with solar r-process ratios. In particular,
Sr exhibits wide variations not shared by
Ba (Ryan et al. 1991, 1996), suggesting that Sr at least does not come from
a universal process. Even in CS~22892-052, the light neutron-capture elements
Sr ($Z$=38) to Cd ($Z$=48) do not match the solar r-process pattern
(Sneden et al. 2000), providing more evidence of different processes affecting 
light and heavy neutron-capture elements.

Although the material from which Pop II stars form is not expected to contain 
significant s-process contributions, some stars including some subgiants are 
greatly enriched in carbon and s-process elements 
(Norris, Ryan, \& Beers 1997a; Hill et al. 2000). 
These are believed to be binary companions of initially
more-massive donor stars which have evolved through the thermally pulsing AGB 
phase and transferred material enriched in C and s-process elements onto the 
lower mass, longer lived secondary now observed. 

In this report, we present the results of two studies of neutron-capture
elements in Pop II stars. The first is for a C-rich, s-process-rich
star in which many heavy elements including lead have been measured. 
The second is a Sr-rich Pop II giant, whose abundance patterns we use to
search for the nucleosynthesis process responsible.

\section{s-process production in Population II stars}

Although the s-process contributes little to GCE during the 
formation of the halo, the measurement of s-process production by 
metal-deficient objects is important for two reasons. Firstly, s-process yields
depend strongly on metallicity (e.g. Busso et al. 1999), so we 
need to know what this dependence is if we are to 
compute accurately the onset of s-process contributions to GCE.
Secondly, the site of the main s-processing is believed to be thermally pulsing
AGB stars. The yields of such stars indicate the conditions in their interiors,
providing information on their structure and evolution.  

The s-process contribution to the interstellar medium that became locked into
newly formed Pop II stars was insignificant, but AGB stars in binaries may 
transfer material to companions which, if of lower mass and longer lived, may 
preserve the Pop II s-process products on their enriched surfaces.
Two high-proper-motion stars with [Fe/H] $\simeq$ $-2.7$ and huge carbon 
excesses [C/Fe] $\simeq$ 2.0 were investigated by Norris et al. (1997a) ---
LP~625-44 and LP~706-7. By virtue of their proper motions, we infer that they
are subgiants (assuming they are bound to the Galaxy).
The spectroscopic analysis also suggests subgiant (rather than giant) 
evolutionary states. In addition to the large carbon over-abundances, the stars
have high s-process abundances, with
[Ba/Fe] = 2.6 and 2.0 respectively. Norris et al. sought to explain the stars
by the mass transfer scenarios described above, but were troubled by
the lack of definite radial velocity variations over a time span of several 
years, which weakened the appeal to a binary
mass transfer process as being responsible.

Aoki et al. (2000) have completed a new investigation of LP~625-44 using 
higher S/N spectra. Definite radial velocity variations have been detected, with
a velocity range $\Delta v ~\ge$ 10 km~s$^{-1}$ over a period $T~\ge$~12~years. 
This implies a wide orbit separation $a~>$ 5~AU, consistent with the wide
separations found for some CH-subgiants where mass transfer from an AGB donor 
is believed to occur by wind accretion rather than Roche-lobe overflow 
(Han et al. 1995).
A more detailed element analysis has also become possible,
the highlight of which was the detection of lead via the 4057~\AA\ absorption. 
Comparison of the spectrum of LP~625-44 with that of CS~22957-027,
a metal-poor star ([Fe/H] = $-3.4$) having a strong $^{12}$C {\it and} $^{13}$C
over-abundances [C/Fe] = 2.2 but no enhancement of neutron-capture elements 
(Norris, Ryan, \& Beers 1997b), verifies that the absorption line is not due
to an unrecognised $^{12}$CH or $^{13}$CH line.

The Pb abundance was measurable in this object because of its overall high
abundance of s-process elements. However, it was the {\it low}
value of the [Pb/Ba] ratio ([Pb/Ba] $\simeq$ 0) that caused initial surprise.
It is well established theoretically and observationally that the ratio of
heavy s-process elements around Ba, to lighter s-process elements around
Sr, [hs/ls] (defined by Luck \& Bond 1981), depends strongly on
metallicity. In general, [hs/ls] increases (i.e. Ba becomes over-abundant
relative to Sr) as the metallicity is reduced from solar to 1/10 solar,
due to the smaller number of seed nuclei for the available neutrons
(e.g. Busso et al. 1999, Fig. 16). At still lower metallicities, the models of 
Busso et al. converge to a [hs/ls] value 3 times solar, though at least some 
stars have higher values (e.g. Norris et al. 1997a, Fig. 8).
However, the observed (low) [Pb/Ba] ratio in LP~625-44 emphasises that the
increase in [hs/ls] towards lower metallicities is not necessarily shared by
an analogous [Pb/hs] ratio.

In contrast to the observation of LP~625-44, the `ST' (standard) models of 
Busso et al. (1999, Fig. 12) indicate a high value [Pb/Ba] $\simeq$ 1.5 at 
[Fe/H] = $-2.7$. However, the predicted [Pb/Ba] 
ratio is not uniquely determined by the metallicity of the AGB star but also by 
the profile of the $^{13}$C pocket, which may depend on the mass and rotational
properties of the star. Moreover, Busso et al. note (their Fig. 16) that a 
{\it range} of $^{13}$C-pocket profiles must be adopted in their models to match
the range of [hs/ls] values observed in higher metallicity stars. 
Consequently, the `ST' model is only one of several
possible solutions. A wider range of models (see Figure~1)
show the sensitivity of the elements to the extent of the $^{13}$C
pocket, and indicate for a 1.5~M$_\odot$ star
at [Fe/H] = $-2.7$ that while [Ba/Fe] may be
up to 0.5 dex above the `ST' model prediction, [Pb/Fe] can range over 
$\sim 1.0$~dex above and below the `ST' value. That is, even though the
Ba enhancement can be predicted within a narrow range, the [Pb/Ba] ratio
depends sensitively on the $^{13}$C profile.
The low [Pb/Ba] abundance measured in
LP~625-44, which initially seemed to challenge the published `ST' models,
may still be within the range of metal-poor AGB models.
According to the
models shown in Figure 1, this requires that only a very weak
burning of $^{13}$C occurred in the intershell zone, with an efficiency
lower by a factor of 20 than in the ST case.

The $^{13}$C profile is not well constrained, but
measurements of many s-process elements in stars like LP~625-44
and LP~706-7, each enriched by a single AGB star,
present the opportunity to search amongst the models for the best fit and
hence to identify the most likely companion/donor. It is hoped that such 
studies will
advance modelling of AGB stellar structure and nucleosynthesis by providing
constraints on actual element yields for AGB stars of low [Fe/H].

\begin{figure}
% \vspace{5cm}   % empty space for the figure (5cm in the given example)
\begin{center}
\leavevmode
\epsfysize=98mm
% \epsfbox{torino00_f1b.eps}
\epsfbox{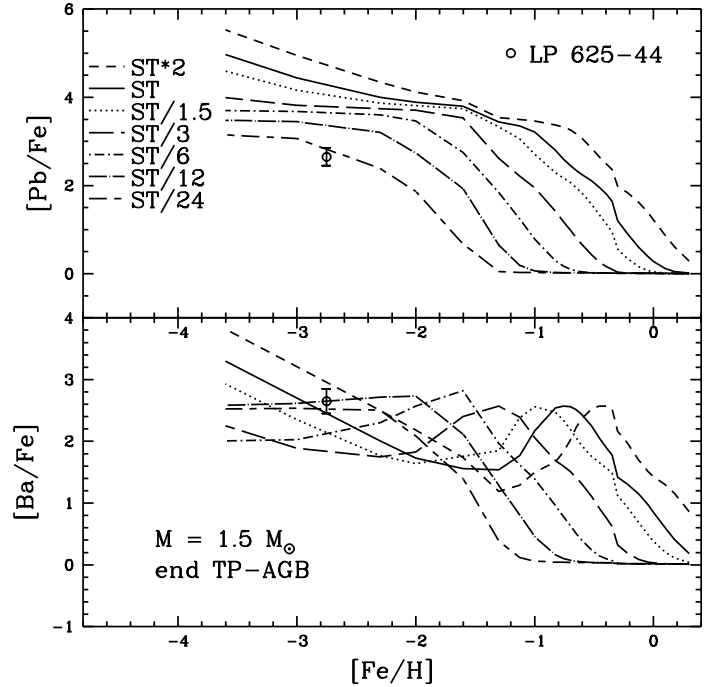}
\end{center}
\caption[h]{Metallicity-dependence of [Pb/Fe] and [Ba/Fe] nucleosynthesis in a
1.5~M$_\odot$ AGB star, for different $^{13}$C profiles. 
{\it Solid curve}: `ST' (standard) model shown by Busso et al. (1999).
{\it Other curves}: other normalisations of the $^{13}$C pocket.
At the metallicity of LP~625-44 (shown),  the [Ba/Fe] ratio is only weakly 
sensitive to the $^{13}$C pocket, but the [Pb/Fe] value is very sensitive to 
this choice of parameter.}
\end{figure}

A separate series of models for neutron-capture nucleosynthesis is the 
1--3.5~M$_\odot$ metal-poor stars ([Fe/H] $<$ $-2.5$) considered by
Fujimoto, Ikeda, \& Iben (2000; their Case II'). The helium flash in these 
metal-poor models drives convection that penetrates into the hydrogen-rich 
envelope and produces C-rich, N-rich material there. These stars later mix 
C-rich and s-process-rich 
material to their surfaces. It remains to be seen whether the s-process element
abundances that such objects would produce are consistent with those observed,
and therefore whether this particular mechanism could have affected 
LP~625-44's mass donor.

\section{Non-universal r-process abundances}

Pop II neutron-capture elements are believed to originate in the r-process,
except in stars enriched after formation by s-process material 
as discussed in the previous section.\footnote{Note, however,
Magain's (1995) finding of s-process-like isotope ratios in HD~140283.}
Although Ba and heavier elements
seem to consistently fit the solar r-process pattern where such data exist
(e.g. Gilroy et al. 1988; McWilliam et al. 1995; Cowan et al. 1995; 
Sneden et al. 1996), at lower atomic numbers a wide variety of element ratios
exists (Ryan et al. 1991, 1996), and even in CS~22892-052 they appear not to
match the solar r-process (Sneden et al. 2000). Figure~2 shows that at
[Fe/H] $^<_\sim$ $-3$, a large range of [Sr/Fe] values exists, 
spread over 2~dex, whereas [Ba/Fe] occupies a much smaller range.

\begin{figure}
% \vspace{5cm}   % empty space for the figure (5cm in the given example)
\begin{center}
\leavevmode
\epsfysize=79mm
% \epsfbox[32 407 526 689]{aat00a_f1.ps}
\epsfbox[32 408 526 689]{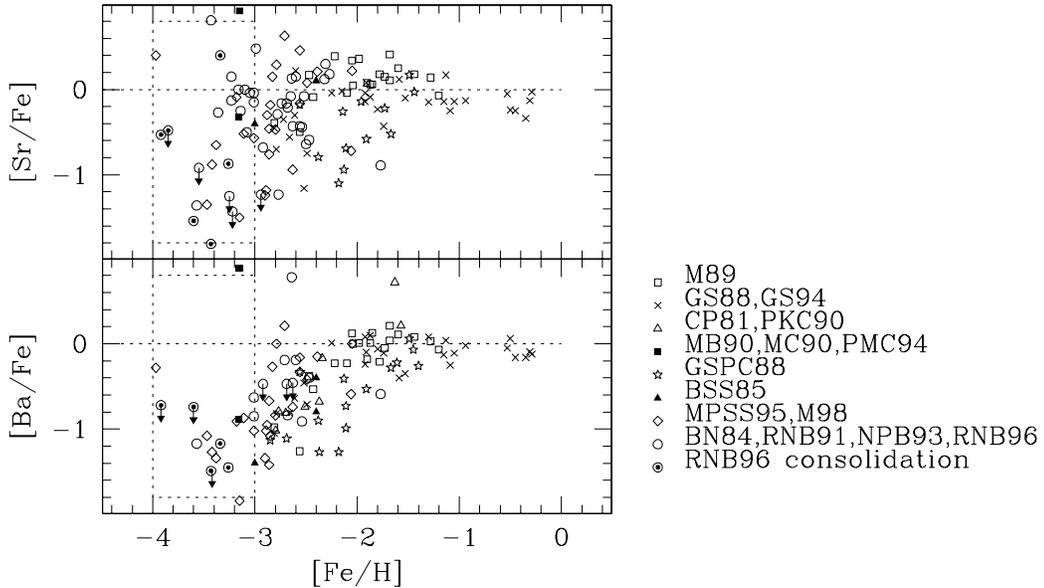}
\end{center}
\caption[h]{Abundances of Sr and Ba in halo stars. The dashed box outlines the
same region of [X/Fe] and [Fe/H] in each plot. Whereas [Sr/Fe] exhibits a range
by a factor of more than 100, [Ba/Fe] has a well-behaved trend towards lower
values at the lowest [Fe/H]. (Ryan 2000)}
\end{figure}

As the lower envelopes to the [Sr/Fe] and [Ba/Fe] values seem similar, and 
because of the observational and theoretical arguments that Ba is 
an r-process product in these stars, it seemed reasonable to regard the lower 
Sr abundances as also produced by the r-process. We then seek processes that 
produce an excess of Sr but not Ba in some metal-poor stars, giving rise 
to the stars with the higher [Sr/Fe] values. It also seemed more natural
to hypothesise an additional source of Sr rather than a means of destroying Ba.
Since the wide ranging [Sr/Fe] ratios are seen in both dwarfs and
giants, we consider processes in the objects that enriched the 
material from which these stars formed, rather than 
self-enrichment or other processes affecting them after their formation.

% We sought high S/N, high resolving power spectra of elements around Sr, to 
We sought high S/N, high resolving power spectra, to 
verify previous abundance estimates and to seek additional elements, in 
particular Zn ($Z$=30) which lies just above the Fe-peak and can be produced 
% along with Sr in some processes. Means of Zn production include 
along with Sr. Means of Zn production include 
(e.g. Woosley \& Weaver 1995, \S4.7):
(1) the $\alpha$-rich freeze-out from explosive Si-burning, where the density
is low enough to avoid complete nuclear 
statistical equilibrium [NSE] with the result that excess $\alpha$-particles 
combine with iron-peak nuclei to extend the distribution up to Zn, also 
producing much Ni (Sneden \& Crocker 1988; Arnett 1995);
and (2) the weak s-process in core-He-burning stars which also produces
Sr, Y, and Zr but not heavier species (Prantzos, Hashimoto, \& Nomoto 1990).
However, quantitative models of GCE do not currently reproduce zinc
observations well, and we cannot be certain that we understand its 
production.
For example, Nomoto et al's (1997) SN II calculations averaged
over the IMF for 10--50~M$_\odot$ stars (their Fig. 8) underproduces the solar
Zn abundance by $^>_\sim$~1~dex, as foreshadowed by their earlier
comments (Thielemann, Nomoto, \& Hashimoto 1996) that additional 
Zn is produced in the s-process and possibly in SN Ia.
The GCE calculations of Timmes, Woosley \& Weaver (1995, Fig. 35) similarly fail
to match the observations,
though Woosley \& Weaver (1995, \S4.7) specifically note that they regard 
the probable source of the dominant solar isotope, $^{64}$Zn, to be an extreme 
$\alpha$-rich freeze-out which was {\it not} included in their study.

Blake et al. (2000a,b) report results for CS~22897-008, a giant selected for its
high [Sr/Fe] and low [Ba/Fe] values.\footnote{It has only a moderate carbon 
abundance ([C/Fe] = 0.34; McWilliam et al. 1995) so is unlikely to be related to
the C-rich, s-process-rich stars investigated by Norris et al. (1997a) or to
the r-process-rich object CS~22892-052 (Sneden et al. 1995).} Through \'echelle
spectra with S/N = 47--129, and $\lambda/\Delta \lambda$ = 50000, 
earlier Sr, Y, Zr, and Ba measurements have been improved, and the
first detection of Zn in this object was obtained.
Abundances, calculated using WIDTH6 (Kurucz \& Furenlid 1979)
and a Bell et al. (1976) model for parameters 
$T_{\rm eff}$/log~$g$/[Fe/H]/$\xi$ 
= 4850/1.8/$-3$/1.9, are shown in Figure~3.
They confirm that Sr, Y, Zr, and also Zn ([Zn/Fe] = +0.7) are more abundant in 
this star than most others of the same [Fe/H].

\begin{figure}
% \vspace{5cm}   % empty space for the figure (5cm in the given example)
\begin{center}
\leavevmode
\epsfysize=75mm
% \epsfbox[32 407 526 689]{abA3rot.ps}
% \epsfbox{abA3rotB.ps}
% \epsfbox[32 220 526 660]{abA3rotBBW.ps}
% \epsfbox{abA3rotBBW.ps}
\epsfbox[32 050 526 380]{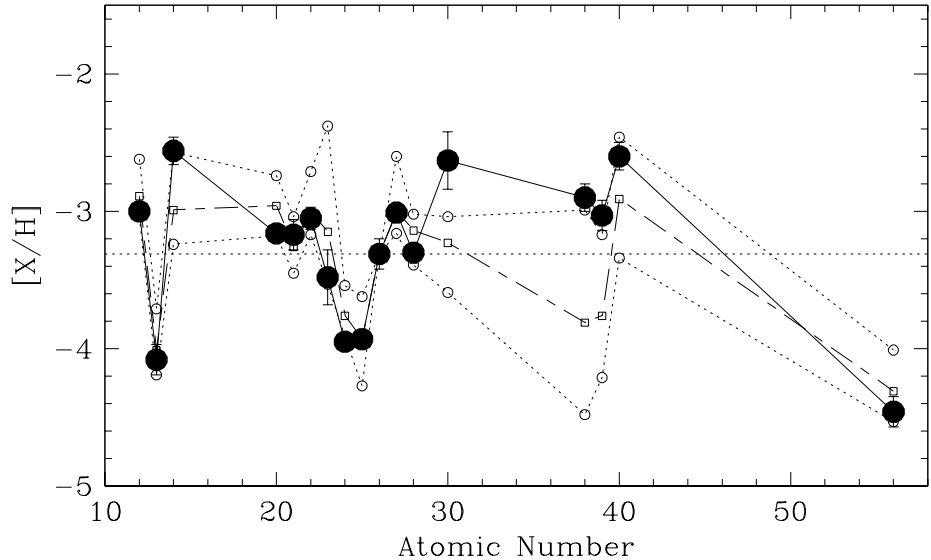}
\end{center}
\caption[h]{[X/H] for a range of elements in CS~22897-008 (solid symbols).
The horizontal line passes through Fe; displacements from this line
give [X/Fe]. The squares (connected by the dashed line) give [X/H] 
values typical of other stars at the same [Fe/H],
% as CS~22897-008,
while open circles approximate the upper and lower quartiles of the spread.
(For V, Zn, and Zr, where few similarly metal-poor stars have been observed, 
comparison values are based on higher metallicity objects.)}
\end{figure}

There is an additional caveat for Zn
in addition to the usual uncertainties in the atmospheric parameters and 
models.  
The `normal' value of [Zn/Fe] in stars at [Fe/H]~=~$-3.3$ is not
well known, due to the paucity of measurements in such 
metal-poor objects. Primas et al. (2000) show only one object with
[Fe/H]~$<$~$-3.0$, for which they give {\it preliminary} values 
[Fe/H]~=~$-3.5$, [Zn/Fe]~=~0.45. It will be interesting to see whether their
final sample reveals an increasing overabundance of [Zn/Fe] in 
more-metal-poor stars, as their preliminary sample suggests and as Johnson 
(2000) also identifies.

Several nucleosynthetic sites that may produce the 
overabundances shown in Figure~3 will be discussed below, with quantitative 
calculations presented elsewhere (Blake et al. 2000b).
Whatever produces Zn, Sr, Y, and Zr abundantly in CS~22897-008
and other stars must avoid producing much Ba. 
If this is a neutron capture process, then a
small exposure, or a small number of neutrons per seed, is required. 

\subsection{The weak s-process}

The weak s-process occurs during He- and C-burning 
in massive stars ($M$~$^>_\sim$~10~M$_\odot$), producing only the
first s-process peak, the $N$~=~50 closed shell at Sr-Y-Zr, but not the 
$N$~=~82 peak at Ba-La (e.g. Prantzos et al. 1990). 
It differs in both site and yield from the main s-process
associated with thermally-pulsing AGB stars (e.g. Busso et al. 1999).
Prantzos et al. show (their Fig.~7) that sufficiently massive stars will 
produce a strong overabundance of elements from Zn to Zr, as would be needed
to explain the elements enhanced in CS~22897-008 (Figure~3).
Furthermore, this nucleosynthetic component would be more obvious in the 
ejecta of massive stars, 
% which is 
the mass range appropriate to the enrichment
of the most metal-poor stars. However, 
% a substantial drawback for our application is that 
it depends strongly on metallicity,
% of the site, 
and is not expected to be significant in metal-poor
objects
% , least of all those 
with [Fe/H]~$<$~$-3$, due to the greater
relative importance of {\it primary} neutron poisons like $^{12}$C, $^{16}$O,
and $^{20}$Ne (Prantzos et al. 1990; Raiteri et al. 1991a,b, 1992; 
Travaglio et al. 1996). 

\subsection{A weak r-processes}

In view of the difficulties faced by the weak-s-process, it is natural to
hypothesise an analogous weak-r-process (e.g. Ishimaru \& Wanajo 2000) that 
produces only the lighter neutron-capture species abundantly, but at the high 
neutron densities associated with a primary r-process. 
Wasserburg, Busso \& Gallino (1996) were prompted by the $^{129}$I/$^{127}$I and
$^{182}$Hf/$^{180}$Hf ratios in the early solar system to hypothesise two 
distinct r-process sites to 
explain light ($A$~$\le$~130) and heavy ($A$~$>$~130) nuclei. 
Qian, Vogel \& Wasserburg (1998) and Wasserburg \& Qian (2000) developed the
idea further, and associated light r-element production with lower frequency, 
higher yield events above a new neutron-star remnant, while heavy r-element 
production occurred in higher frequency but lower yield events associated with 
the formation of black hole remnants.

\subsection{The $\alpha$-process}

The $\alpha$-process --- an $\alpha$-rich freeze-out 
(e.g. Woosley, Arnett, \& Clayton 1973)
from explosive Si-burning
which avoids complete nuclear statistical
equilibrium in the low density, high temperature neutrino bubble
above a 1--10-second-old neutron star --- is capable of producing not only
excess Zn but also Sr-Y-Zr (Woosley \& Hoffman 1992; 
Witti, Janka \& Takahashi 1994; Woosley et al. 1994).
One of the challenges of explaining the excess Sr in the stars at 
[Fe/H]~$<$~$-3$ is that it appears in some but not others, and 
with varying magnitude. That is, the data exhibit a large spread in
[Sr/Fe], not simply a bimodal distribution.\footnote{The bimodality in the 
[Ba/Fe] distribution discussed by Ryan et al. (1996) was eliminated 
(see Figure~3) by McWilliam's (1998) revision of the Ba abundances for his 
sample.} The existence of a wide spread, covering $^>_\sim$~2~dex, suggests 
that the efficacy of whatever mechanism is responsible 
is strongly dependent upon some key factor(s).
Takahashi et al. (1994) show that a 
reduction in the density of the neutrino bubble by a factor of 5 would allow 
the normal r-process to proceed rather than the $\alpha$-process. 
With such a small variation in density required to toggle between the
Sr-Y-Zr-producing $\alpha$-process and the r-process, 
it may be possible to switch between normal and excessive production of these
species with just the natural variation in density. This is an attractive
feature of the $\alpha$-process possibility. 
% It remains to
% be shown why its effect should be obvious only for the
% most metal-poor stars.

\subsection{Neutron-star mergers}

Freiburghaus et al. (1999) have shown that neutron-star mergers could be an
important site for the production of r-process elements with $A$~$>$~130
in the Galaxy. The $Y_{\rm e}$ range explored in their work finds 
underproduction of $A$~$<$~130 species, which fits qualitatively the data of
Sneden et al. (2000) for CS~22892-052. However, this result is in the opposite 
sense to that found for the stars we are investigating, which in contrast have 
high Sr and low Ba abundances. That is, while neutron-star mergers may prove 
responsible for much of the r-processing in the Galaxy, they do not, by current
calculations, describe the nucleosynthesis that has produced high Sr abundances
in the most metal-poor stars.

\section{Concluding remarks}

These stars emphasise that element analyses of 
neutron-capture elements in very metal-poor stars can constrain the 
nucleosynthesis processes and sites that have operated in the Galaxy. This will
provide a greater degree of realism in the models and hence to our computations
of the chemical evolution of the Galaxy. In the case of very metal-poor stars, 
we hope to constrain models of sites that may no longer be observed directly.

Caution is also needed to avoid overinterpreting the observations. In many 
discussions of heavy-element production, the element abundances in CS~22892-052 
are cited as if they were a definitive record of early Galactic nucleosynthesis.
While these data are exquisite, and many of the recent advances in our 
understanding of neutron-capture elements have been based on this star's 
composition, people should bear in mind that this is a very {\it a}typical star.
It has a carbon overabundance of a factor of ten, and is one of few 
metal-poor stars showing strong r-process overabundances, again by a factor of
$^>_\sim$~10 (depending on exactly which element is measured). It is precisely 
because it is such an {\it unusual} star that it has been able to be studied in
such detail. We must be cautious that we do not infer that the typical products
of nucleosynthesis in the early epoch of {\it Galaxy} are those seen in such an 
unusual and atypical single object.

\acknowledgements
It is a pleasure to acknowledge many
valuable discussions with 
% Dr C. Travaglio.
Drs Y. Ishimaru, C. Travaglio, and S. Wanajo.
% Financial support for S.G.R.'s attendance at the workshop was provided by 
% the Universit\'a di Torino and The Open University.

\end{document}